\documentclass[11pt]{article}
\usepackage{amssymb,amsmath}

\def\wig{_{{\rm\scriptstyle w}}}
\def\R{{\mathbb{R}}}      % real numbers
\def\C{{\mathbb{C}}}      % complex numbers
\def\H{\mathcal{H}}       % Hilbert spaces

\begin{document}
\centerline{\bf On the notion of quantum Lyapunov exponent.}

\bigskip
\centerline{ M.F. Kondratieva$ ^*$ and T.A. Osborn$ ^{**}$}

\bigskip

\noindent {\small {\it *) Department of Mathematics and Statistics,  Memorial
University of Newfoundland, St. John's, NL, Canada} A1C 5S7}

\noindent {\small {\it **) Department of Physics and Astronomy, University of
Manitoba, Winnipeg, MB, Canada} R3T 2N4}

\bigskip

\noindent
{\bf Abstract}.
Classical chaos refers to the property of  trajectories to diverge
exponentially  as time $t\to \infty$.
It is characterized by a positive Lyapunov
exponent.

There are many different descriptions of quantum chaos. The one related to the
notion of generalized (quantum) Lyapunov exponent is based either on
qualitative physical considerations or on the so-called  symplectic tomography
map \cite{Mendes, MendMan}.

The purpose of this note is to show how the definition of quantum Lyapunov
exponent naturally arises in the framework of the Moyal phase space formulation
of quantum mechanics \cite{MQM}, and is based on the notions of quantum
trajectories and the family of quantizers \cite{Strat}.  The role of the
Heisenberg uncertainty principle in the statement of the criteria for quantum
chaos is made explicit.

\bigskip
{\bf 1. Introduction}.\bigskip

Irregular behaviour of classical dynamical systems arising from deterministic
time evolution without any external randomness and stochasticity -- the so
called deterministic chaos -- manifests itself as an extremely sensitive
dependence on the initial conditions, which makes unstable the long-time
prediction of the dynamics.

In such a system, a positive Lyapunov exponent is a quantitative measure of the
infinite time exponential separation of neighbouring orbits.

In detail, let the system have a form $\dot x=F(x)$, where $x=(q,p)$ is
$d$-dimensional vector from the system's phase space. Denote by $x(t, x_0)$ its
solution with initial point at $x_0$ and $t\in (0,\infty)$. Then the Lyapunov
exponent is given by
$$
\lambda =\lim_{t\to\infty}\frac 1 t \ln \frac{||\delta x(t)||}{||\delta
x(0)||}\,.
$$
Here $||\cdot||$ represents the $d$-dimensional Euclidian norm, and
 $||\delta x(0)||$ is initial infinitesimal deviation from $x_0$, $||\delta x(t)||=||
x(t,x_0+\delta x(0))- x(t,x_0)||$ is deviation from $x(t,x_0)$ at time $t$.
 In the limit
$||\delta x(0)||\to 0$ one obtains
\begin{equation}
\label{lambda} \lambda_v = \lim_{t\to\infty}\frac 1 t \ln \left[ (v\cdot
\nabla_{x_0})\, x(t,{x_0}) \right],
\end{equation}
where $v\in R^d$ is a unit vector in the direction of the initial
displacement, $\delta x(0)$, and $\nabla_{x_0}$ is gradient w.r.t.
the initial point $x_0$.

To extend this notion to quantum mechanics it is natural to use
its phase space formulation where quantum observables on the
Hilbert space $\H = L^2(\R^n,\C)$ are represented by functions on
the phase space, called symbols. Consider Weyl symbol $A(x)$,
obtained from the operator $\hat A$ s.t. $\hat A\psi(q') \equiv\int  \,\langle
q'|\hat A | q''\rangle\langle q''|\psi\rangle\,dq''$, $\psi\in \H$, by formula
\begin{equation}
\label{weyls} A(x)= 2^n \int e^{2ips/\hbar}\langle q-s|\hat A|q+s\rangle  ds\,
\equiv[\widehat A\,]\wig (x)\,.
\end{equation}
The dimension of the phase space is even, $d=2n$.

In this case quantum mechanical mean value
\begin{equation}
\label{mean1} \langle\widehat A\rangle_\rho ={\rm Tr}\, \widehat
A\,\hat \rho=\int \langle q'|\hat A|q''\rangle \langle q''|\hat \rho|q'\rangle\, dq'dq''
\end{equation}
can be written in a form analogous to  classical statistical mechanics
\begin{equation}
\label{mean2} \langle \widehat A\rangle_\rho= \int A(x)\, W(x)\,  dx\,,
\end{equation}
which gives mean value of $A$ in the state $\rho(x)=[\hat \rho]\wig (x)$. Here
$W$ is the Wigner distribution function $W(x)=h^{-n}\rho(x)$.

One  drawback of this approach, relative to the classical statistical density,
is that the $W(x)$ is not everywhere nonnegative so is not a conventional
probability density.

To circumvent this difficulty another representation can be considered in which
a new density is defined as the Radon transform \cite{Gelfand} of the Wigner
function $W(x)$. It takes only non-negative values and becomes straightforward
analog of the classical statistical density. In \cite{Bertrand} it was shown
that the tomographic representation of quantum mechanics based on Radon
transform is an alternative to the Weyl--Wigner formalism.

Due to their similar nature in quantum and classical cases the tomographic
distributions were used in \cite{MendMan} to define a quantum Lyapunov
exponent. In detail, let $U(t)=exp\,[-it\widehat H/\hbar]$ be the Schr\"odinger
unitary evolution operator, and $\widehat X(t)=U^+(t)\, \hat x\, U(t)$ be time
evolution of the quantum coordinate operators $\hat x=(\hat p,\hat q)$. The
formula for quantum Lyapunov exponent turned out to be
\begin{equation}
\label{lam} \Lambda_{v} =\lim_{t\to\infty}\frac 1 t \ln || \langle\widehat X(t)\rangle_{\rho_0^v} ||\,,
\end{equation}
where average is taken with respect to a special initial singular
density $\hat \rho_0^v$, the kernel of which has the form
\begin{equation}
\label{ker} \langle q'|\hat \rho_0^v |q''\rangle= e^{ip_0(q'-q'')/\hbar}\left(
(v_1\nabla) +\frac {iv_2(q'-q'')}\hbar \right)
\delta\left(q_0-\frac{q'+q''}2\right).
\end{equation}
Here parameters $(q_0,p_0)$ correspond to the initial point $x_0$
in the phase space, and vector $v=(v_1,v_2)\in \R^d$ defines the
direction of the initial deviation from $x_0$. The authors of
\cite{MendMan} stress  special role of the tomographic
distributions for obtaining these formulas, in particular the one
for the initial density (\ref{ker}). In \cite{Mendes} similar
formula is obtained from qualitative considerations but the choice
of the initial density appears rather {\it ad hoc} in the quantum
mechanical setting. As  it is shown in \cite{Mendes}\cite{MendMan},
quantum Lyapunov exponent (\ref{lam},\ref{ker}) helps to classify different types of quantum complexity.
There are examples where exponential rate of growth for the trace $\langle\widehat X(t)\rangle_{\rho_0^v}={\rm Tr}\widehat X(t)\widehat\rho_0^v $
of position and momentum observables
starting from the singular initial density matrix (\ref{ker}) was found in quantum mechanics. In many cases when quantum mechanics has
damping effect on the classical chaos and the rate of growth  for the trace is milder than exponential,
the notion of quantum sensitive dependence was used instead.

In this paper we derive representations of the quantum Lyapunov
index from the Weyl--Stratonovich quantizer \cite{Strat}. Using notion of
quantum trajectory \cite{MQM}, i.e.  the symbol of operator  $\widehat X(t)$ ,  we rewrite the formula for quantum Lyapunov exponent
in the form identical  to the classical definition
(\ref{lambda})
(see formula (\ref{lambd})), replacing classical trajectory $x(t,x_0) $ with quantum trajectory $X(x_0,t;\hbar)$. In this form it becomes transparent that
 the definition respects the
correspondence principle: in the limit $\hbar\to 0$ definition of quantum Lyapunov exponent transforms into the classical one.
In contrast to the classical trajectory, the Heisenberg uncertainty principle prevents the quantum trajectory to be interpreted as a
measurable physical value despite the fact that it can be viewed as a limit of a sequence of quantum means.

\bigskip
{\bf 2. Quantum Phase Space. Weyl--Stratonovich quantizer.}\bigskip

Development of phase space formulation of quantum mechanics has a long history,
but it still is of interest due to extensive study of possible generalizations
to the case of non-Abelian gauge theory with support on a Riemannian manifold.

In 1932 Wigner introduced his quasi-probability distribution associated with
the wave function $\psi(q)$
\begin{equation}
\label{WF} W(x)= \left({\frac {2}{h}}\right)^n \int
e^{2ips/\hbar}\,\psi^+(q-s)\,\psi(q+s)\,ds\,,\quad h=2\pi\hbar.
\end{equation}

Further developing Groenewold's ideas published in 1946, Moyal gave in his
paper of 1949 statistical interpretation of the Wigner's formula as a Fourier
inverse of the expectation value of the Heisenberg translation operator
\begin{equation}
\label{Exp} T(y)\equiv\exp(-2iJy\cdot \hat x/\hbar)\,,  \qquad T(x)^+\, \hat
x\, T(x)=\hat x-2xI.
\end{equation}
namely
$$
W(x)={\left(\frac 2{h}\right)}^{2n} \int\, e^{2ix\cdot Jy
/\hbar}\langle\psi|T(y)|\psi\rangle\,dy\,.
$$ where $J$ denotes the Poisson matrix $\left[\begin{array}{cc}
0&I\\-I&0\end{array}\right]$. He also showed that the Wigner rule
(\ref{WF}) of getting the phase space function from the operator
$|\psi\rangle \langle \psi|$ (and $\widehat A$) is inverse to the
Weyl quantization rule,
\begin{equation}
\label{weyl} \langle q'|\widehat A | q''\rangle=h^{-n} \int
e^{ip(q'-q'')/\hbar} A\left(\frac{q'+q''}2,p\right) \, dp\,.
\end{equation}

A significant step further was the introduction \cite{Strat,Grosmann,Royer} of
the family of unitary operators $\Delta (x)$ labelled by points of the phase
space. These operators define both quantization (\ref{weyl}) and dequantization
(\ref{weyls}) rules, which establishes a unitary isomorphism between symbols
and operators,
\begin{equation}
\label{both} A(x)=2^n\,{\rm Tr}\, \Delta(x)\, \widehat A\,, \quad
\widehat A =\int A(x)\,\Delta(x)\,d^*x \,. \quad
\end{equation}
For this reason the operators $\Delta(x)$ are called quantizers.
Above $d^*x = (\pi\hbar)^{-n}dx$ denotes a dimensionless phase
space measure.

The fundamental nature of quantizers also reveals itself in the
fact that they define the noncommutative product $*$ for the phase
space functions
\begin{eqnarray}  &(A*B)(x)&= \int A(y)\,B(z)\,
K(x,y,z) \, dy\, dz, \\\ &K(x,y,z) &=
\frac{2^n}{(\pi\hbar)^{2n}}\,{\rm Tr}\,
\Delta(x)\,\Delta(y)\,\Delta(z)\,.
\end{eqnarray}
Other useful properties \cite{GH78} of quantizers are:

\bigskip \noindent 1. $\Delta(x)=\Delta(x)^+=\Delta(x)^{-1}$. Thus
$\Delta(x)^2=I $ and $||\Delta(x)||=1\,$;

\smallskip
\noindent 2. ${\rm Tr}\, \Delta(x)= 2^{-n}$\,;

\smallskip
\noindent 3. ${\rm Tr}\, [\Delta(x)\,\Delta(x')]= (\pi\hbar/2)^n\, \delta(x-
x')\,$;

\smallskip
\noindent 4. $\Delta(x)=\int e^{2ix\cdot Jy/\hbar}\,  T(y)\, d^*y\,$;

\smallskip
\noindent 5. $\rho(x)= 2^n \langle\psi| \Delta(x) |\psi \rangle\,$;

\smallskip
\noindent 6. $\Delta(0)\, \hat x\,\Delta(0)=-\hat x\,$;

\smallskip
\noindent 7. $\Delta(x)= T(x/2)^+ \Delta(0)\, T(x/2)\,$;

\smallskip
\noindent 8. $\int  \Delta(x)\,d^*x\,=I$.
\bigskip

Many authors (see e.g. \cite{Gracia}) use quantizers as a
fundamental object defining the deformation quantization
introduced in 1978 by Bayen et al \cite{BBF78}.

There is one observation useful for the purpose of our note. In
view of the 1st equation of (\ref{both}) Property 3 tells us that
the symbol of the quantizer is $[\Delta(x)]\wig
(x')=(h/2)^n\delta(x-x')$, and by (\ref{weyl}) that its kernel is
\begin{equation}
\label{kernel} \langle q'|\Delta(x) | q''\rangle= 2^{-n}
 e^{ip(q'-q'')/\hbar}\,
\delta\left(q-\frac{q'+q''}2\right)\,.
\end{equation}

    An attractive feature of quantum phase space is computation of
the trace of an operator and pairs of operators. For an operators
$\widehat A, \widehat B$ with symbols $A(x),B(x)$ one has
\begin{eqnarray}
\label{tr1}{\rm Tr}\, \widehat A &=& \frac 1{h^n} \int A(x)\, dx\,, \\
\label{tr2}\, {\rm Tr}\, \widehat A\, \widehat B &=& \frac 1{h^n}
\int A*B(x)\, dx = \frac 1{h^n} \int A(x)\,B(x)\, dx\,.
\end{eqnarray}
Identity (\ref{tr1}) follows from (\ref{both}) and Property 2. The
removal of the $*$ operation in (\ref{tr2}) is a consequence of
Properties 3 and 8.\bigskip

{\bf  3. Quantum means and symbols.}
\bigskip

Mean value of a quantum observable given by operator $\widehat A$
in a unit normalized quantum state $\psi(q)$ is
\begin{equation}
\label{meanpsi} \langle\psi | \widehat A |\psi\rangle ={\rm
Tr}\,\widehat A\, \hat \rho\,, \quad
\hat\rho=|\psi\rangle\langle\psi |\,.
\end{equation}
This is an example of formula (\ref{mean1}) for the pure state
density. Its phase space form is (\ref{mean2}) which is a special
case of the trace identity (\ref{tr2}).

Let us consider a family of Gaussian states  localized near
$q=q_0\in R^n$ with width $\sqrt\hbar$
$$
\psi_\hbar (q;q_0,p_0) =\frac 1{(\pi \hbar)^{n/4}} \exp
\left(-\frac{(q-q_0)^2}{2\hbar} + \frac i \hbar p_0(q-q_0)\right).
$$
These states are all unit normalized, $||\psi_\hbar
(q_0,p_0)||=1.$ By formula (\ref{WF}) corresponding normalized
Wigner function has the form
$$
W_\hbar(x; q_0,p_0)= \frac 1{(\pi\hbar)^n} \exp \left(
-\frac{(q-q_0)^2+(p-p_0)^2} \hbar\right).
$$
Two important remarks should be made about this function.\smallskip

{\it Remark 1}. Function $W_{\hbar}(x;x_0)$ is  positive  and thus can be
interpreted as the classical statistical density. \smallskip

{\it Remark 2}. In  view of the following formula
\begin{equation}
\label{limite} \lim_{\hbar \to 0} W_\hbar(x;
x_0)=\delta(x-x_0)\end{equation} it  is evident that the symbol of
the quantizer $\Delta(x)$ is proportional to the limit of the
sequence of Wigner functions $W_\hbar(x; x_0)$ of the localized
Gaussian states as their width $\sqrt\hbar$ goes to 0.

Thus the $\hbar$-independent symbol $A(x_0)$ of an operator $\widehat A$
evaluated at point $x_0$ of the phase space appears as the mean value
calculated with respect to the localized Gaussian state with parameters
$x_0=(q_0,p_0)$ in the limit when the width of the state goes to zero
$$
\lim_{\hbar \to 0} \langle\psi_\hbar (q;x_0) |\widehat
A|\psi_\hbar (q;x_0)\rangle = \lim_{\hbar \to 0}\int
A(x)W_\hbar(x; x_0)\,dx = A(x_0).
$$

Note that this limiting process takes one outside the
framework of quantum expectation values. The norm
$||\Delta(x)||=1$ implies that for any pure state $\hat \rho =
|\psi\rangle \langle\psi|$
\begin{equation}\label{bound}
|\rho(x)|\leq 2^n \quad {\rm and} \quad |W(x)|\le \left(\frac 2
h\right)^n.
\end{equation}
This means that the quantizer or any singular symbol such as
$\delta(x-x_0)$ does not correspond to a pure state. The
global bound (\ref{bound}) forces $\rho(x)$ to be distributed in
phase space without large peaks and is an evident consequence of
the uncertainty principle.

Let quantum operator $\widehat A$ have Weyl symbol $A(x)$, and
 time dependent operator $\widehat A(t)=U^+(t)\widehat AU(t)= A(\widehat X(t))$
has Weyl symbol $A(x,t;\hbar)$ such that $A(x,0;0)=A(x)$. Consider now the
following expression
$$
\langle A\rangle(t, x_0; \hbar,\varepsilon)=\int
W_{\varepsilon}(x;x_0) A(x,t;\hbar)\, dx
$$
as a function of two small parameters $\hbar $ and  $\varepsilon$. The
following table shows meaning of the expression if one or both of the
parameters are 0.

\bigskip
\begin{tabular}{|c||c|c|}
\hline
$\langle A\rangle(t, x_0; \hbar,\varepsilon)$&$\hbar \to 0$&$\hbar\ne 0$\\
\hline
\hline
$\varepsilon\to 0$&classical observable
&symbol $A(x_0,t;\hbar)$ of \\
&$A(x(t,x_0))$ &quantum observable $\widehat A(t)$\\
\hline
$\varepsilon\ne 0$&classical statistical mean
&quantum mechanical \\
&$\int dy \, W_{\varepsilon}(y;x_0) A(x(t,y)) $&mean value\\
\hline
\end{tabular}

\bigskip Note that function $W_{\varepsilon}$ for $\varepsilon\ne
\hbar$ does not correspond to a pure state $\psi(q)$, but rather
is a density matrix.

\bigskip
{\bf 4. Classical and quantum Lyapunov exponents.}\bigskip

From the table we see that the symbol $A(x_0,t;\hbar)$ is in fact quantum analog of
the classical value $A(x(t,x_0))$. This is due to Egorov Theorem which
states
\begin{equation}
\label{Ego}
\lim_{\hbar \to 0 }A(x,t;\hbar) =A(x(t,x_0)),
\end{equation}
where $x(t,x_0)$ is classical trajectory.

Thus the quantum analog of $(v\cdot \nabla_{x_0}) A(x(t,x_0))$ will be
$(v\cdot \nabla_{x_0}) A(x_0,t;\hbar)$.

From the 1st formula of (\ref{both}) one has
$$
A(x_0,t;\hbar)=2^n \int  \langle q'|\Delta(x_0) | q''\rangle\langle
q''|\widehat A(t)| q'\rangle\, dq'' dq'\,.
$$
Here we see that whole information about the point $x_0$ of phase space is
contained in the quantizer kernel $\langle q'|\Delta(x_0) | q''\rangle$. Thus
the derivative $(v\cdot \nabla_{x_0})$ will only effect this part of the
formula
$$
(v\cdot \nabla_{x_0}) A({x_0},t;\hbar)= 2^n\int\left[(v\cdot
\nabla_{x_0})\langle q'|\Delta({x_0}) | q''\rangle \right]\,
\langle q''|\widehat A(t)| q'\rangle\, dq'' dq' \,.
$$

From (\ref{kernel}) we obtain
$$
2^n(v\cdot \nabla_{x_0})\,\langle q'|\Delta({x_0}) |
q''\rangle=\langle q'|\hat\rho_0^v|q''\rangle\,,
$$
where $\langle q'|\hat\rho_0^v|q''\rangle$ is given by $(\ref{ker})$. So far we
have
\begin{equation}
\label{avad} (v\cdot \nabla_{x_0}) A({x_0},t;\hbar)= \int\langle
q'|\hat\rho_0^v|q''\rangle \langle q''|\widehat A(t)| q'\rangle\,
dq'' dq' = \langle \widehat A(t)\rangle_{\rho_0^v}
\end{equation}

Now, let $A(x)$ be a vector function.
Introduce
\begin{equation}
\label{lamq} \Lambda^{A}_v (x_0;\hbar)=\lim_{t\to \infty}\frac 1 t
\ln || (v\cdot \nabla_{x_0}) A(x_0,t;\hbar)||\,.
\end{equation}
Using (\ref{avad}) we can also write
\begin{equation}
\label{lamq1} \Lambda^{A}_v (x_0;\hbar)=\lim_{t\to \infty}\frac 1 t \ln ||
\langle\widehat A(t)\rangle_{\rho_0^v}||.
\end{equation}

Taking limit $\hbar \to 0$ in (\ref{lamq}) we get
\begin{equation}
\label{lamc}
\Lambda^{A}_v(x_0;0)=\lim_{t\to \infty}\frac 1 t \ln ||
(v\cdot \nabla_{x_0}) A(x(t,x_0))||.
\end{equation}
This formula is well defined for a wide class of vector functions $ A(x)$. To
get the classical Lyapunov exponent one must make  however the special choice
$A(x)= x$. Then formula (\ref{lamc}) becomes exactly (\ref{lambda}).

In the quantum case symbol $A(x)=x$ defines operator $\hat x$ which makes it
possible to talk about the {\it quantum trajectory} \cite{MQM} defined as the
symbol of $\widehat X(t)=U^+(t)\hat x U(t)$
$$
X(x_0,t;\hbar)=
\lim_{\varepsilon \to 0}{\rm Tr }
\widehat X(t) \widehat \rho_\varepsilon=
 \lim_{\varepsilon \to 0}\int
 X(x,t;\hbar)W_\varepsilon(x; x_0))\,dx\,.
$$

Then formula (\ref{lamq1}) gives us definition of the quantum Lyapunov
exponent, which coincides with (\ref{lam}), (\ref{ker}).

In view of (\ref{Ego}) $\lim_{\hbar\to 0} X(x_0,t;\hbar)=x(t,x_0)$, and
the notion of quantum trajectory and (\ref{lamq}), (\ref{lamq1}) allows also to write (\ref{lam}) in a form similar to
(\ref{lambda})
\begin{equation}
\label{lambd}
\Lambda_v=\Lambda^{X}_v (x_0;\hbar)=
\lim_{t\to\infty}\frac 1 t \ln \left\{ (v\cdot \nabla_{x_0})\,
X(x_0, t;\hbar) \right\}.
\end{equation}

%Then formula  becomes universal for both cases, the classical and
%the quantum.

\bigskip
{\bf 5. Radon transform and tomographic procedure}.\bigskip

Let us make a few comments on the approach undertaken in [2]. To
simplify exposition of some formulas and their geometrical meaning
we assume in this section that $n=1$, i.e $q\in \R^1$, $x\in
\R^2$.

Although the integral of the Wigner function for the density
matrix $|\psi\rangle \langle \psi|$ is always 1, typically there
are regions in phase space where $W(x)$ is negative.  This
behavior results from the spectral expansion of the quantizer.
Property 1 shows that the spectrum of $\Delta(x)$ is $\pm 1$ for
all $x$. In detail
\begin{equation}\label{sp1}
\Delta(x) = P_+(x) - P_{-}(x)\,,\qquad I = P_+(x) + P_{-}(x)
\end{equation}
where $P_{\pm}(x)$ are the spectral projectors for the eigenvalues
$\pm 1$. If $x=0$, then $\Delta(0)$ is parity operator on $\H$ and
$P_{\pm}(0)$ are the corresponding even and odd projectors. When
$x\neq 0$ then $P_{\pm}(x)$ are the Heisenberg translates of these
operators, namely $T(x/2)^+ \,P_{\pm}(0) T(x/2)$. Applying
(\ref{sp1}) to $W(x)$ gives the Royer expansion
\begin{equation*}
W(x) = \frac 2h \left[ ||P_+(x)\psi||^2 - ||P_{-}(x)\psi||^2
\right]\,.
\end{equation*}
So whenever $||P_+(x)\psi|| < ||P_{-}(x)\psi||$ then $W(x)$ is
negative. This makes it impossible to interpret it as a classical
statistical density.

Nevertheless the two physically important {\it projections} of the $W(x)$,
namely
$$
\int  W(q,p)\,dp=|\psi(q)|^2\,,\quad \int W(q,p)\,dq=
|\widetilde\psi(p)|^2
$$
are positive and form corresponding marginal distributions. Here $\widetilde\psi(p)=h^{-n/2}\int\psi(q)e^{-ipq/\hbar}dq$ is the
wave function in the momentum representation.

One may consider a family of {\it projections} with respect to all directions,
in the phase space, not only the two given above. In this way we obtain Radon
transform of $W(x)$
$$
RW(Q,\xi,\eta)=\int W(q,p)\,\delta(Q-q \xi-p\eta)\,dq\,dp\,.
$$
Here $Q-q \xi-p\eta=0$ is an equation of line in the phase space.
Rewriting the above formula as
\begin{eqnarray*}
RW(Q,\xi,\eta)&=&\frac 1{2\pi}\int W(q,p)e^{-ik(Q-q
\xi-p\eta)}\,dk dq\,dp
\\    &=&\frac 1{2\pi}\int dk e^{-ikQ} \int dq\,dp\,W(q,p) e^{i( kq \xi+k p\eta)}
\end{eqnarray*}
one can see two remarkable things. First, the RHS can be viewed as
a inverse Fourier of a characteristic function $\int
W(q,p)\exp\{i( kq \xi+k p\eta)\}\,dq\,dp\,$ and thus is
non-negative and represents a marginal distribution. Second, Radon
transform is a composition of 2D-Fourier transform and 1D-inverse
Fourier transform, thus it is invertible. The inverse Radon
transform can be written as
$$
W(q,p)=\int dQ d\xi d\eta\, RW(Q,\xi,\eta) e^{ i (Q-q
\xi-p\eta)}.
$$
This means that knowing  {\it projections} in all directions one can
reconstruct $W(x)$ from
them by the inverse Radon transform.
This property is widely used in computer tomography.

Due to invertibility of Radon transform the tomographic representation is
equivalent to Weyl symbols representation. In particular the formula for the
quantum mean takes form (\ref{mean2})
$$
\langle\widehat A\rangle_\rho = \int dx\, A(x)W(x) =\int dQ d\xi d\eta
 \tilde A(\xi,\eta) RW(Q,\xi,\eta) e^{ iQ},
$$
where $\tilde A (\xi,\eta)$ is inverse Fourier of $A(q,p)$, and
$RW(Q,\xi,\eta)$ is the tomographic distribution corresponding to $W(x)$.

Any of these representations finally lead to the same formula for the quantum
Lyapunov exponent.

\bigskip
This research was supported by grants to M.F.K. and T.A.O. from the Natural Sciences and Engineering Research Council of Canada.


\smallskip

\begin{thebibliography}{1}

\bibitem{Mendes}
 R Vilela Mendes, {\it Sensitive dependence in quantum systems: some
examples and results} Phys Lett A 171(1992) 253-258;

\bibitem{MendMan} V I Man'ko, R Vilela Mendes, {\it Lyapunov exponent in quantum mechanics.
A phase space approach}
Physica D 145 (2000) 330-348;

\bibitem{MQM} T A Osborn and F H Molzahn, {\it Moyal Quantum Mechanics: The semiclassical Heisenberg
dynamics} Ann. Phys. (N.Y.) 241 (1995) 79-127.

\bibitem{Strat} R.L. Stratonovich, {\it On distributions in representation space} Sov Phys JETP 4 (1957) 891-898.

\bibitem{Gelfand} I.M. Gelfand, M.I. Graev and Ya. Vilenkin,
{\it Generalied functions} vol 5;
I.M. Gelfand, S. G. Gindikin,  M.I. Graev, {\it Selected Problems in Integral
Geometry} Dobrosvet, Moscow 2000 (in Russian)


\bibitem{Bertrand} J. Bertrand, P.Bertrand, {\it Tomographic procedure for
constructing phase space representations.
in The Physics of Phase Space} Lecture Notes in Physics v.278, Springer 1986.


\bibitem{Grosmann} A. Grosmann, {\it Parity operator and quantization
of $\delta$-functions}, Commun
Math Phys., v. 48(1976), 191-193.

\bibitem{Royer} A. Royer, {\it  Wigner function as the expecttion value of
a parity operator},
Phys Rew. A, v. 15 no 2 (1977), 449-450.

\bibitem{GH78}
A.~Grossmann and P.~Huguenin.
\newblock {\it Group-theoretical aspects of the {W}igner--{W}eyl
isomorphism.}
\newblock  Helvetica Physica Acta, 51 (1978) 252--261.

\bibitem{Gracia} J.M. Gracia-Bondia, {\it Generalized Moyal Quantization on
Homogeneous Symplectic Spaces} Contemp. Math v. 134 (1992), 93-113.

\bibitem{BBF78}
{F.~Bayen, M.~Flato, C.~Fronsdal, A.~Lichnerowicz, and
D.~Sternheimer}.
\newblock {\it Deformation theory and quantization.}
\newblock  Ann. Phys. (N.Y.), 111 (1978) 61--110, .


\end{thebibliography}
\end{document}